\documentclass{iau} 
\usepackage{graphicx}

\title[Torki et al] 
{From evolved Long-Period-Variable stars to the evolution of M31}

\author[]   
{Maryam Torki$^1$, Mahdieh Navabi$^1$,	Atefeh Javadi$^1$, Elham Saremi$^1$, Jacco Th. van Loon$^2$
	and Sepideh Ghaziasgar$^1$ }

\affiliation{$^1$School of Astronomy, Institute for Research in Fundamental Sciences (IPM), \\ Tehran, \\[\affilskip]
	$^2$Lennard-Jones Labratories, Keele University, ST5 5BG, UK }

\pubyear{2015}
\volume{xxx}  
\jname{Title of your IAU Symposium}
\editors{A.C. Editor, B.D. Editor \& C.E. Editor, eds.}
\begin{document}

\maketitle

\begin{abstract}
One of the ways to understand the genesis and evolution of the universe is to know how galaxies have formed and evolved. In this regard, the study of star formation history (SFH) plays an important role in the accurate understanding of galaxies. In this paper, we used long-period variable stars (LPVs) to estimate the SFH in the Andromeda galaxy (M31). These cool stars reach their peak luminosity in the final stage of their evolution; their birth mass is directly related to their luminosity. Therefore, we construct the mass function and the star formation history using stellar evolution models.
\keywords{stars: AGB and post-AGB - stars: luminosity function, mass function - galaxies: evolution - galaxies: formation - galaxies: individual: M31 - galaxies: stellar content - galaxies: structure}
\end{abstract}

\firstsection 
\section{Introduction}

 Andromeda, as the closest spiral giant galaxy, offers us a unique opportunity to know how its various components have formed and evolved. M31 is located at 785± 25 kpc, $(m-M)_o=24.47$ mag (McConnachie et al. 2004) and has low foreground reddening E(B-V)=0.06 mag. Our approach to investigating the star formation history (SFH) is based on employing long-period variable stars (LPVs), which we have successfully applied in other galaxies in the Local Group, such as M33 (Javadi et al. 2011a; 2011b; 2017), Magellanic Clouds (Rezaeikh et al. 2014), NGC147 and NGC185 (Golshan et al. 2011), IC1613 (Hashemi et al. 2019), Andromeda VII (Navabi et al. 2020; 2021) and Andromeda I (Saremi et al. 2021). Asymptotic giant branch (AGB) and red supergiants (RSGs) are basic pillars in our research.

\section{Data and Technique of Star Formation}
In this paper, we used the catalogue of LPVs in M31 from Mould et al. (2004) to determine star formation (Torki et al. 2019). It should be noted that their results included near-infrared photometry of almost 2000 variables, most of which were AGBs. We used the Padova stellar evolutionary models (Marigo et al. 2017) to obtain SFH and assuming that the metallicity is constant over time, we were able to calculate the mass, age, and pulsation duration of the stars. The SFH is described by the star formation rate (SFR) and conveys the concept of how much gas has become a star each year in the past. In fact, SFR, $\xi$ (in M$_\odot$ yr$^{-1}$) is a function of time and estimated by:

\begin{equation}
	\xi(t) = \frac{\int_{\rm min}^{\rm max}f_{\rm IMF}(m)m\ dm}
	{\int_{m(t)}^{m(t+dt)}f_{\rm IMF}(m)\ dm}\ \frac{dn^\prime(t)}{\delta t}.
	\label{eq:eq8}
	\end{equation}

 \begin{figure}[tb]
	\begin{center}
		\includegraphics[width=6in]{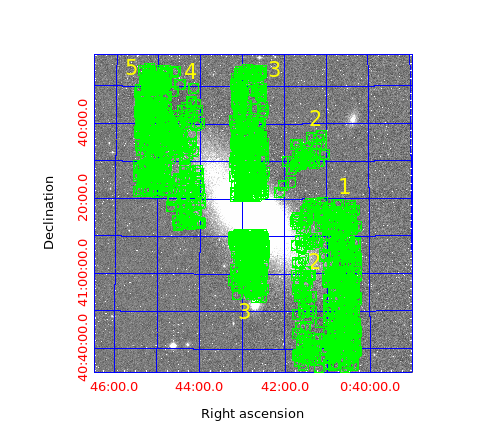} 
		\caption{The 2MASS image of the M31 galaxy. The green dots indicate the LPVs from Mould et al (2004).}
		\label{fig1}
	\end{center}
\end{figure}

\begin{figure}[tb]
	\begin{center}
		\includegraphics[width=4.7in]{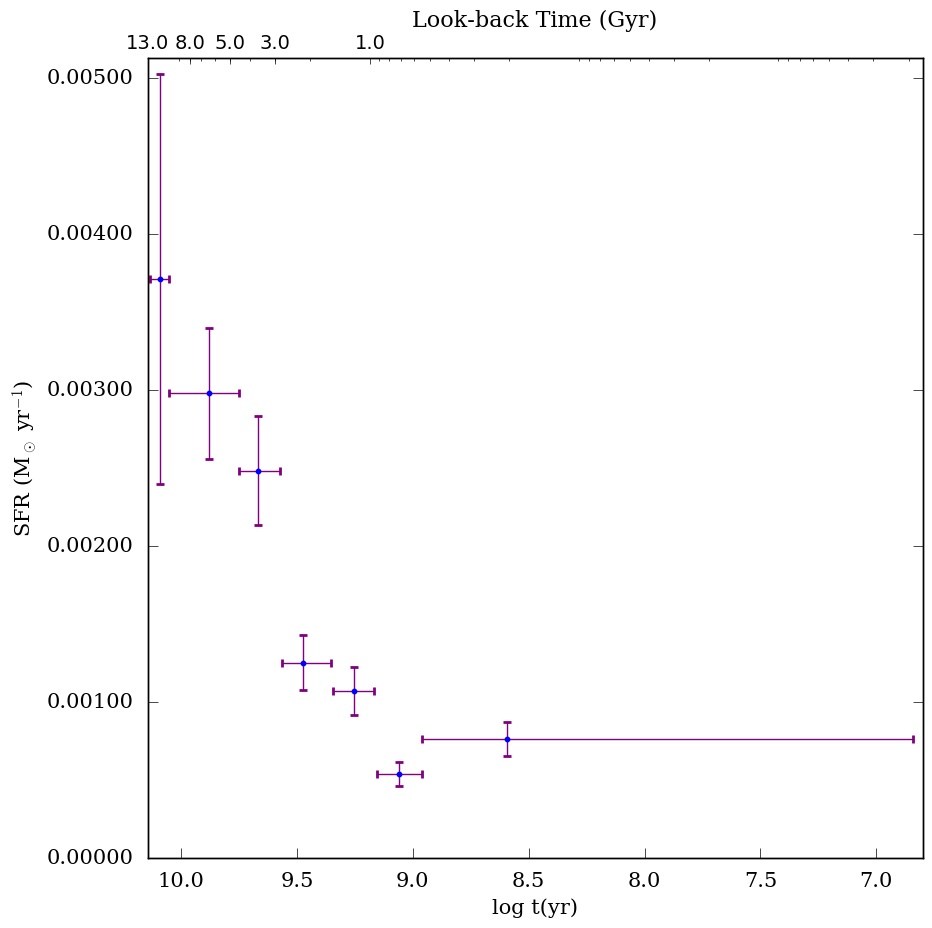} 
		\caption{SFH of the M31 galaxy for constant metallicity of Z = 0.008 based on data in box 1 of Figure 1. The horizontal “error bars” show the spread in age within each bin and the statistical errors are presented in the vertical “error bars”.}
		\label{fig1}
	\end{center}
\end{figure}

 where the m is mass, $f_{\rm IMF}$ is the Initial Mass Function (IMF) (Kroupa 2001). The minimum and maximum mass in Kroupa's IMF are considered 0.02 and 200 M$_\odot$ respectively. In the above equation, it is assumed that stars with mass between $m(t)$ and $m(t+dt)$ (t is look back time) are LPVs at the present time, and $\delta t$ is the duration of the star evolutionary phase, which the number of variables observed ($dn^\prime(t)$) in intervals $t$  and $t+dt$ depends on.

\section{RESULTS} 
The LPVs are scattered in five rectangular boxes (Figure 1). In this paper, we estimated the SFH of the M31 galaxy during the broad time interval from 
11.8 Gyr (log $t$ = 10.07) to 400 Myr (log $t$ = 8.59) ago in box 1, which its dimensions are 1.49 kpc$\times$13.09 kpc. The SFR as a function of look-back time in M31 is shown in Figure 2. The horizontal “error bars” correspond to the start and end time for which the SFR was calculated in that bin. We see a peak of star formation at 11.8 Gyr ago; after that, the star formation begins to decrease until the present-day. \\

\end{document}